\documentclass[final,3p,times]{elsarticle}

\usepackage{amssymb}    
\usepackage{amsmath}    
\usepackage{lipsum}  
\usepackage{graphicx}   
\usepackage{subcaption}
\usepackage{accents}    
\usepackage{array}      
\usepackage{color}      
\usepackage{hyperref}

\hypersetup{
  pdftitle={A Fully Resolved Multiphysics Model of Gastric Peristalsis and Bolus Emptying in the Upper Gastrointestinal Tract},
  pdfauthor={Shashank Acharya}
}

\journal{ArXiv}

\begin{document}

\begin{frontmatter}

\title{A Fully Resolved Multiphysics Model of Gastric Peristalsis and Bolus Emptying in the Upper Gastrointestinal Tract}

\author{Shashank Acharya\fnref{LB1}}
\author{Sourav Halder\fnref{LB2}}
\author{Wenjun Kou\fnref{LB3}}
\author{Peter J. Kahrilas\fnref{LB3}}
\author{John E. Pandolfino\fnref{LB3}}
\author{Neelesh A. Patankar\corref{cpau}\fnref{LB1,LB2}}
\cortext[cpau]{Corresponding author}
\ead{n-patankar@northwestern.edu}

\address[LB1]{Department of Mechanical Engineering, Northwestern University, Evanston, IL 60208, USA}
\address[LB2]{Theoretical and Applied Mechanics Program, Northwestern University, Evanston, IL 60208, USA}
\address[LB3]{Division of Gastroenterology and Hepatology, Feinberg School of Medicine, Northwestern University, Chicago, IL 60611, USA}

\begin{abstract}

Over the past few decades, \textit{in silico} modeling of organ systems has
significantly furthered our understanding of their physiology and
biomechanical function. In this work, we present a detailed numerical model of
the upper gastrointestinal (GI) tract that not only accounts for the fiber
architecture of the muscle walls, but also the multiphasic components they
help transport during normal digestive function. Construction details for 3D
models of representative stomach geometry are presented along with a simple
strategy for assigning circular and longitudinal muscle fiber orientations for
each layer. Based on our previous work that created a fully resolved model of
esophageal peristalsis, we extend the same principles to simulate gastric
peristalsis by systematically activating muscle fibers embedded in the
stomach. Following this, for the first time, we simulate gravity driven bolus
emptying into the stomach due to density differences between ingested contents
and fluid contents of the stomach. This detailed computational model of the
upper gastrointestinal tract provides a foundation on which future models can
be based that seek to investigate the biomechanics of acid reflux and probe
various strategies for gastric bypass surgeries to address the growing problem
of adult obesity.

\end{abstract}


\end{frontmatter}

\section{Introduction and Motivation} \label{sec:intro}

The digestive system is one of the main organ systems of the body and is
responsible for ingestion, transport, breakdown and absorption of food
necessary for normal body function. In spite of its relative importance,
little attention has been paid to \textit{in silico} modeling of this organ
system compared to extensive numerical and experimental modeling efforts of
the cardiovascular, respiratory and skeletal systems of the human body. The
modeling of biological organ systems can further our understanding of the role
of biomechanical processes in these systems and help differentiate between
physiological and pathological conditions leading to better treatment planning
and outcomes. Up to this point, computational investigations of
gastrointestinal biomechanics have focused on the fluid and solid problems
separately \cite{Ferrua2010, Ferrua2011, Pal2004, Peirlinck2017, Yassi2009}.
Flow in the stomach was generated by fully specifying the motion of gastric
walls with time \cite{Ferrua2011, Pal2004} without accounting for the
interaction between the elastic stomach wall structure and the internal fluid
contents. While this approach reveals important details about flow fields in
the stomach during gastric peristalsis \cite{Imai2013, Pal2007}, it provides
little information on the relationship between pressure fields that drive flow
and the material properties of the surrounding muscular structures. This
information is valuable for probing the integrity of the muscle wall and is
able to relate measurable lumen pressures to material properties of the
esophagus and stomach walls \cite{Gregersen_book_2003}. It must be noted that
significant amount of work has also been done to study the spread and
absorption of pharmocological agents such as tablets and pills in the stomach
\cite{Karkossa2019, Chiang2020, Paixo2018, Hens2019}. Other numerical models
of the gastrointestinal tract have accounted for electrical activity in the
walls and the complex interaction between slow waves generated by the
interstitial cells of Cajal, spike potentials and their overall role in
eliciting contractions of the muscle wall \cite{Brandstaeter2019, Cheng2013, Pullan2004}. 
However, the combined effect of contraction strength and the
structure's material properties on the motion of the confined fluid contents
remains to be explored.

In addition to the coupled fluid-solid phenomena observed during bolus
transport and gastric mixing, it is important to note that the contents of the
digestive system often include more than one homogeneous fluid. Air is often
present in the gastric lumen and there are slight differences in the density
of ingested food and the surrounding gastric fluid (which is similar to
water). These density differences can significantly affect the internal flow
fields and the net effect of the elastic structure on the multiphasic
components must be captured in any reasonably detailed model of the upper GI
tract. In the upright position, gravity is often sufficient to cause fluid
from the esophagus to empty into the stomach \cite{Miki2010, Kahrilas1988,
Pope1997}. This neccisitates the use of two fluid components to model
gravity-driven emptying in the esophagus. In light of the studies summarized
above, we aimed to develop a model that accounted for the complete two-way
fluid-structure interaction (FSI) problem in the stomach along with a
multiphasic approach that accounts for density differences observed during
normal physiologic functioning of the upper GI tract.

\section{Mathematical modeling of FSI and multiphase flow}

\subsection{Description of the Immersed Boundary Method} \label{sec:ib_method_info}

The computational model of the upper gastrointestinal tract presented in this
work is an extension of the model previously developed and validated by Kou et al. to
analyze esophageal peristalsis and bolus transport using the immersed boundary
method \cite{Kou2017_jcp, Kou2017_nmo, Kou2018_bmmb}. Below, we summarize the
mathematical formulation of the immersed boundary finite element (IBFE) method
that was used to model the interaction between the fluid and solid components
in this work. For additional details pertaining to the spatio-temporal
discretization of the governing equations and the Eulerian-Lagrangian
interaction equations, please refer to Refs.~\cite{Kou2017_jcp, Kou2018_bmmb,
EGriffith2017, Nangia2019_dlm, VadalaRoth2020}. In the present setting, the
entire fluid-structure system is modeled using an Eulerian description for
mass and momentum conservation. The structure's deformation and stresses are
modeled using a Lagrangian description. Delta function kernels are used to
exchange information between the Eulerian and Lagrangian subdomains. The
complete set of governing equations can be summarized as follows,


\begin{align}
\frac{\partial\rho\mathbf{u}\left(\mathbf{x},t\right)}{\partial t}+\nabla\cdot\rho\mathbf{u}\left(\mathbf{x},t\right)\mathbf{u}\left(\mathbf{x},t\right) &=
-\nabla p\left(\mathbf{x},t\right)+\nabla\cdot\left[\mu\left(\nabla\mathbf{u}\left(\mathbf{x},t\right)+\nabla\mathbf{u}\left(\mathbf{x},t\right)^{T}\right)\right]+
\rho\mathbf{g}+\mathbf{f}^{e}\left(\mathbf{x},t\right) \label{eq:momentum_conservation} \\
\nabla\cdot\mathbf{u} &= 0 \label{eq:mass_conservation} \\
\mathbf{f}^{e}\left(\mathbf{x},t\right) &=
\int_{U}\mathbf{F}^{e}\left(\mathbf{s},t\right)\delta\left(\mathbf{x}-\boldsymbol{\chi}\left(\mathbf{s},t\right)\right)\mathrm{d}\mathbf{s} \label{eq:force_spreading} \\
\int_{U}\mathbf{F}^{e}\left(\mathbf{s},t\right)\cdot\mathbf{V}\left(\mathbf{s}\right)\mathrm{d}\mathbf{s} &=
-\int_{U}\mathbb{P}^{e}\colon\nabla_{\mathbf{s}}\mathbf{V}\left(\mathbf{s}\right)\mathrm{d}\mathbf{s},\quad\forall\mathbf{V}\left(\mathbf{s}\right) \label{eq:PK1_to_Fe_Vs} \\
\mathbf{U}^{e}\left(\mathbf{s},t\right) &=
\int_{\Omega}\mathbf{u}\left(\mathbf{x},t\right)\delta\left(\mathbf{x}-\boldsymbol{\chi}\right)\mathrm{d}\mathbf{x} \label{eq:velocity_interpolation_int} \\
\int_{U}\frac{\partial\boldsymbol{\chi}}{\partial t}\left(\mathbf{s},t\right)\cdot\mathbf{V}\left(\mathbf{s}\right)\mathrm{d}\mathbf{s} &=
\int_{U}\mathbf{U}^{e}\left(\mathbf{s},t\right)\cdot\mathbf{V}\left(\mathbf{s}\right)\mathrm{d}\mathbf{s},\quad\forall\mathbf{V}\left(\mathbf{s}\right) \label{eq:velocity_interpolation_final} \\
\mathbb{P}^{e} &= \mathcal{G}\left[\boldsymbol{\chi}\left(\cdot,t\right)\right] \label{eq:mat_prop_eqn}
\end{align}

Here, Eqs.~(\ref{eq:momentum_conservation}) and (\ref{eq:mass_conservation})
describe momentum and mass conservation of the entire fluid domain and the
structures immersed within it. Unlike previous versions of the immersed
boundary formulation used in \cite{Kou2017_jcp, Kou2018_bmmb} to model esophageal transport,
Eq.~(\ref{eq:momentum_conservation}) accounts for a variable density and
viscosity that allows for more than one homogeneous fluid to be modeled in the
domain. Eulerian and Lagrangian coordinates are labeled as $\mathbf{x}$ and
$\mathbf{s}$. The combined multiphase fluid-structure system occupies the
space $\Omega \subset \mathbb{R}^3$ and the structure's reference
configuration occupies the space $U$. Velocity and pressure in the Eulerian
description are denoted by $\mathbf{u}$ and $p$. The elastic force densities
in the Eulerian and Lagrangian descriptions are denoted by $\mathbf{f}^e$ and
$\mathbf{F}^e$, respectively. In the Lagrangian frame, the structure's
position, velocity and first Piola-Kirchoff stress are denoted by
$\boldsymbol{\chi}\left(\mathbf{s},t\right)$,
$\partial\boldsymbol{\chi}\left(\mathbf{s},t\right)/\partial t$ and
$\mathbb{P}^{e}$, respectively. The structure's elastic force density is
spread into the surrounding fluid using Eq.~(\ref{eq:force_spreading}) which
then affects fluid flow as seen from the presence of $\mathbf{f}^e$ in
Eq.~(\ref{eq:momentum_conservation}). Elastic force density from the structure
$\mathbf{F}^e$ is computed using $\mathbb{P}^{e}$ and arbitrary Lagrangian test
function $\mathbf{V(\mathbf{s})}$ using a weak form of the principle of
virtual work in Eq.~(\ref{eq:PK1_to_Fe_Vs}). The intermediate Lagrangian
velocity of the structure $\mathbf{U}^e$ is computed by interpolating fluid
velocity using the delta function as shown in
Eq.~(\ref{eq:velocity_interpolation_int}). This intermediate velocity is
projected back into the space defined by $\mathbf{V(\mathbf{s})}$ using
Eq.~(\ref{eq:velocity_interpolation_final}) to obtain the final Lagrangian
velocity of the structure. The reason for this additional step is due to the
fact that the Eulerian equations are solved using the finite difference method
whereas the Lagrangian equations are discretized using finite elements. As
such, the intermediate velocity field must be projected onto the space defined
by the finite element basis functions \cite{VadalaRoth2020}. This final
Lagrangian velocity is then integrated to obtain the updated position of the
structure's nodes. The combination of
Eqs.~(\ref{eq:velocity_interpolation_int}) and
(\ref{eq:velocity_interpolation_final}) ensures that the no-slip boundary
condition is enforced at the fluid-solid boundary. The constitutive model of
the structure is given by Eqn.~(\ref{eq:mat_prop_eqn}) and is used to compute
$\mathbb{P}^{e}$ from the current deformed configuration of the structure.
Thus, Eqs.~(\ref{eq:momentum_conservation})-(\ref{eq:mat_prop_eqn}) illustrate
the mathematical model used to capture the dynamics of the structure due to
the surrounding fluid and vice versa.

\subsection{A brief description of the multiphase flow solver and interface treatment}

As outlined in Sec.~\ref{sec:ib_method_info}, the conservative form of the
Navier-Stokes equations are used to model the overall fluid motion. Density
and viscosity are functions of space and time in
Eq.~(\ref{eq:momentum_conservation}). As the material derivative of $\rho$
remains zero, Eq.~(\ref{eq:mass_conservation}) is equivalent to the general
form of the equation for conservation of mass in a continuum. This
conservative form of the governing equations is discretized using a staggered
grid approach with the traditional description of vector components defined on
cell faces and scalar quantities defined at cell centers. By using mass flux
density alongside the conservative form of the governing equations, the
numerical scheme consistently transports mass and momentum which prevents
any non-physical fluid motions. Special care is taken for the treatment of the
nonlinear advection term which uses a third-order upwind scheme to maintain
monotonicity with higher order accuracy. The overall fluid solver is
second-order accurate for moderate density ratios as commonly observed in
gastrointestinal flows. Complete details of the numerical implementation and
tests for accuracy are provided in \cite{Nangia2019_dlm, Nangia2019_robust}.

The interface between the two fluids is tracked using the level-set method. In
this study, the solid structures are assumed to be of the same phase as the
ambient fluid (water, unless mentioned otherwise). As such, no additional
level set is needed to track the motion of the structure and it remains
neutrally buoyant. A scalar function $\phi(\mathbf{x},t)$ denotes the value of
the level set field in the domain. The variable $\phi$ is a signed distance
function with a value of zero at the interface between the two fluids. Its
value is advected using the fluid velocity field with the equation

\begin{equation}
\frac{\partial\phi}{\partial t}+\nabla\cdot\left(\phi\mathbf{u}\right)=0.
\end{equation}

Fluid material properties are assigned based on the sign of $\phi$ at any
given location. As is well known with traditional usage of level-set methods,
linear advection leads to a gradual loss of the signed distance property of
the $\phi$ field. Reinitialization is periodically conducted by solving the
Eikonal equation $||\nabla\phi||=1$, to recover the signed distance property of
the field. Details on the numerical treatment of this step can be found in
\cite{Nangia2019_dlm}.

\subsection{Software implementation of the mathematical model}

Computations using the governing equations previously described were performed
using the IBFE and Multiphase Flow modules within the IBAMR open source
framework \cite{ibamr_github} which is a distributed-memory parallel
implementation of the IB method written in C++. Equations are solved on a
Cartesian finite-volume grid with adaptive mesh refinement (AMR). Fluid
variables and the AMR strategy are handled using data structures provided by
the SAMRAI library \cite{anderson2013samrai} which uses a patch-based
implementation approach. In addition to the zero level-set interface, the
entire structure is located on the finest level of the finite volume grid.
This ensures that both the fluid-solid interface and the zero-level set are
described using the smallest possible mesh spacing for improved resolution.
The structure is discretized using first-order finite elements and data
related to Lagrangian operations is handled by the libMesh finite-element
library \cite{kirk2006libmesh}. The PETSc solver library \cite{balay2019petsc}
is used to compute numerical solutions for the discretized equation systems.
Computations were performed using clusters PSC Briges and SDSC Comet through
the XSEDE program. Additional computations were conducted on the Northwestern
University high performance computing cluster Quest. All simulations used 96
cores spread over four nodes unless mentioned otherwise.

\begin{figure}[t!]
     \centering
     \begin{subfigure}[b]{0.45\textwidth}
         \centering
         \includegraphics[width=0.95\textwidth]{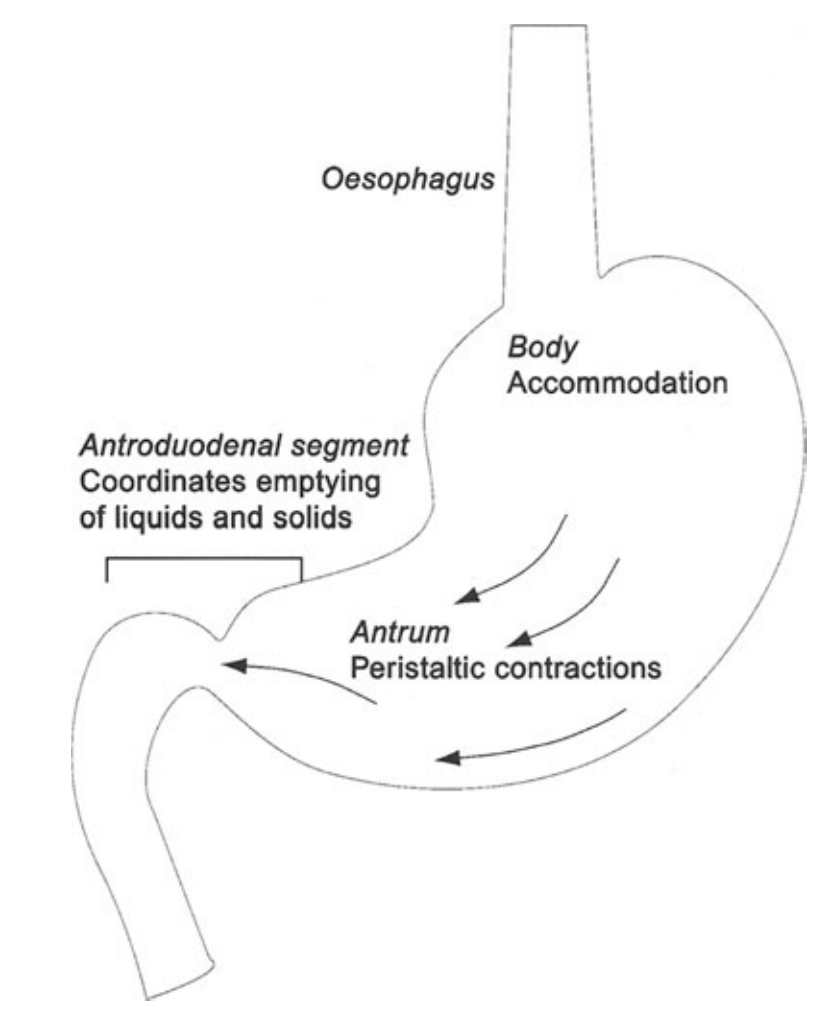}
         \caption{Simplified profile of the stomach showing often observed motility patterns in the proximal and distal stomach \cite{Gregersen_book_2003}.}
         \label{fig:gregersen_stomach}
     \end{subfigure}
     \quad
     \begin{subfigure}[b]{0.45\textwidth}
         \centering
         \includegraphics[width=\textwidth]{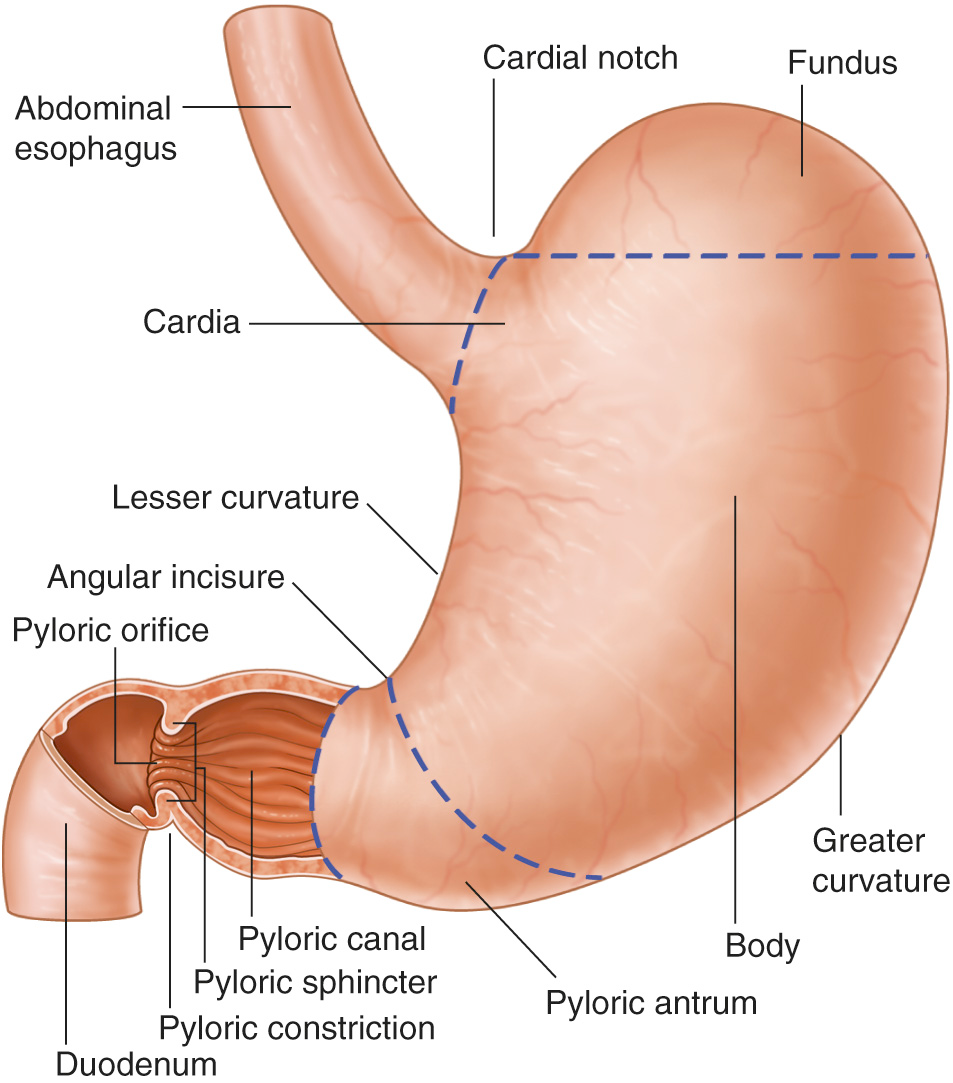}
         \caption{Anatomical components of the stomach and cutaway showing the thicker gastric wall at the pyloric sphincter \cite{drake_grays_2019}.}
         \label{fig:grays_stomach_fibers}
     \end{subfigure}
     \vskip\baselineskip
     \begin{subfigure}[b]{0.45\textwidth}
         \centering
         \includegraphics[width=0.9\textwidth]{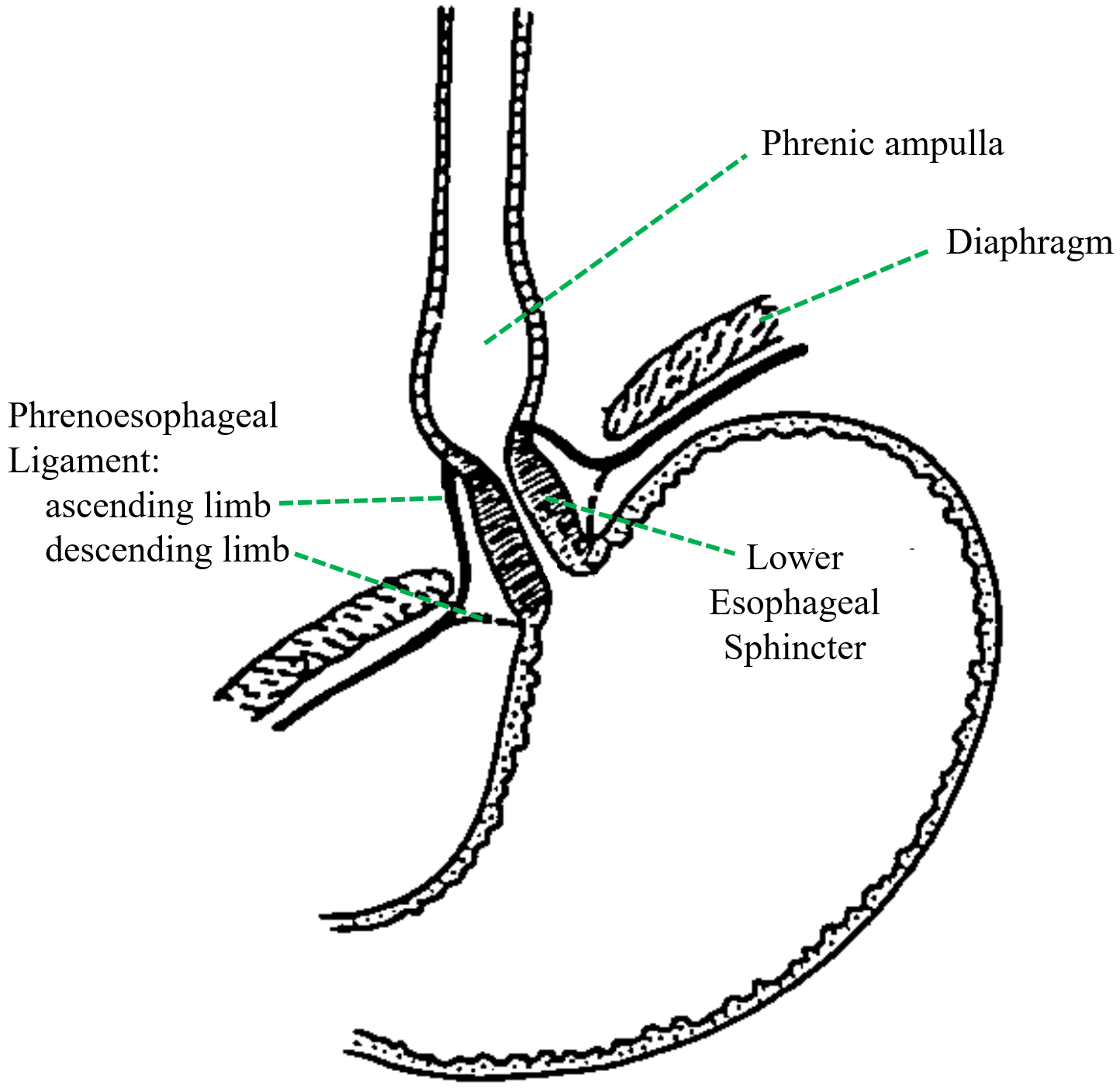}
         \caption{Location of the lower esophageal sphincter (LES) and diaphragm with respect to the stomach volume \cite{Bombeck1966}.}
         \label{fig:bombeck_stomach_PEL}
     \end{subfigure}
     \quad
     \begin{subfigure}[b]{0.45\textwidth}
         \centering
         \includegraphics[width=\textwidth]{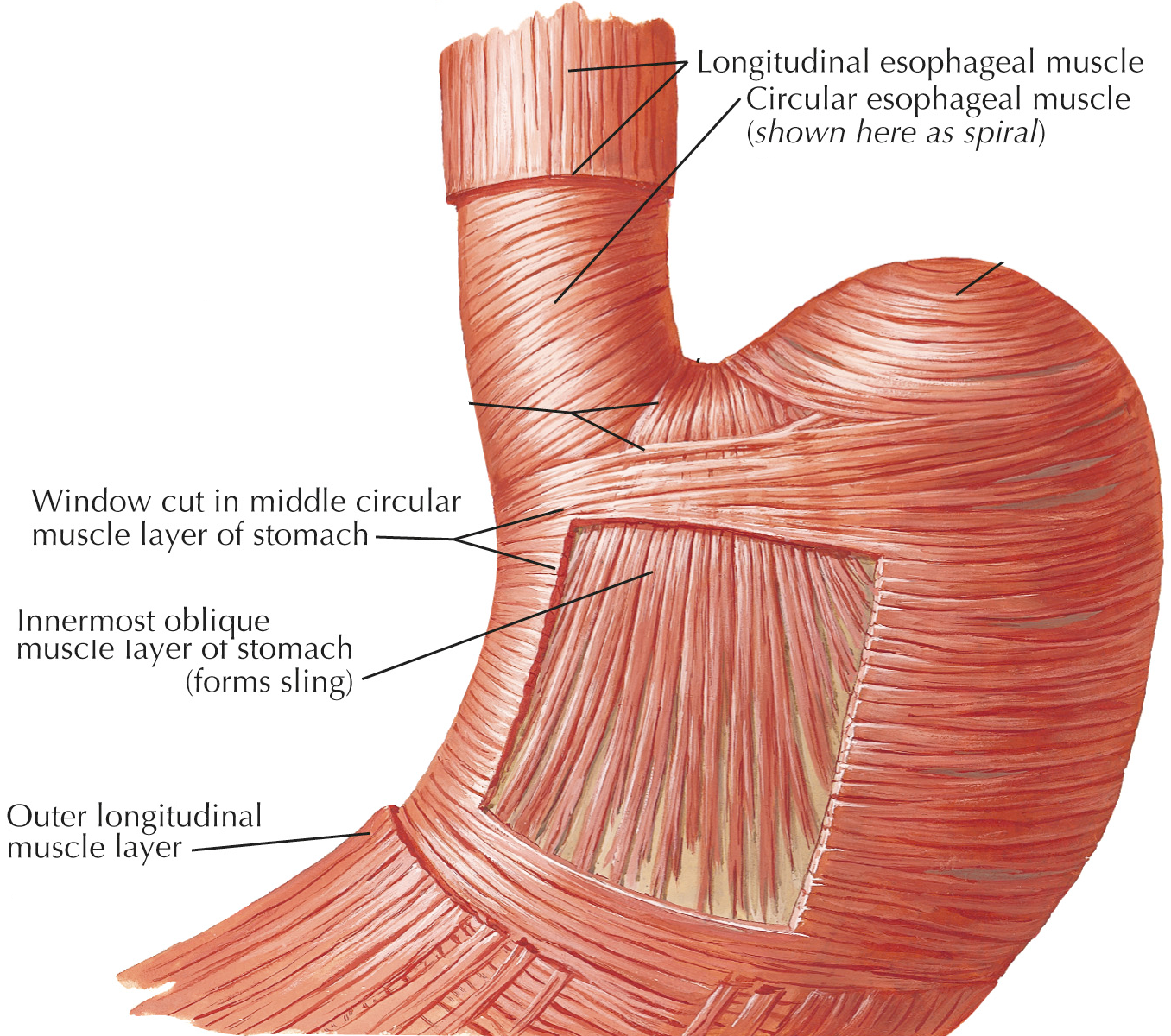}
         \caption{Fiber orientations and muscle layers of the stomach. Muscle fibers from the esophagus continue to form the muscle layers of the stomach \cite{smith_netter_2011}.}
         \label{fig:netter_stomach}
     \end{subfigure}
        \caption{Representative stomach geometry obtained from literature on upper
        gastrointestinal tract anatomy and biomechanics. This collection of
        images highlights the primary geometric features of the stomach along
        with the orientations of fibers in the various muscle layers of the
        stomach walls. The three key geometric features are: 1) The fundus and
        its positioning with respect to the esophagus, 2) the lesser and
        greater curvatures of the stomach, and 3) the pyloric sphincter and
        its approximate muscle thickness.}
        \label{fig:stomach_pictures}
\end{figure}

\section{Construction of representative stomach geometry and associated fiber-architecture}

\subsection{Creating a solid model of the stomach}

With the mathematical model established and the software framework ready for
use, we proceed to describe the construction of the stomach geometry used for
this study. In Fig.~\ref{fig:stomach_pictures}, we show a few images of the
human stomach as commonly seen in literature describing human anatomy. Such
images formed the foundation for the construction of a 3D stomach structure.
The geometry of the esophagus remains identical to the ones used in previous
studies \cite{Kou2017_jcp, Kou2018_bmmb, Kou2017_nmo}. It consists of a thick
walled cylindrical tube with five layers, each having a distinct fiber
architecture that corresponds to various mucosal and muscle layers in a human
esophagus. The stomach geometry however, is far more complex and requires the
construction of some key geometric elements which will then be used to
generate the stomach structure.

\begin{figure}
  \centerline{\includegraphics[width=\textwidth]{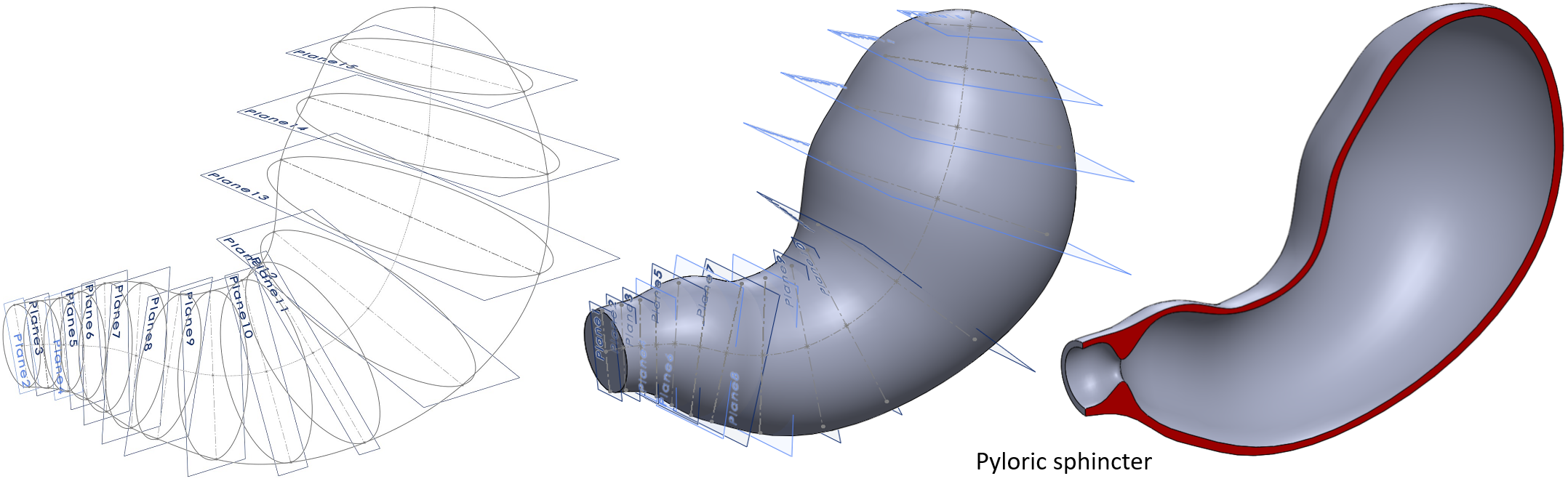}}
  \caption{Solid model of the stomach based on images from
  Fig.~\ref{fig:stomach_pictures}. The image on the left shows all the
  primitive geometric entities that were used to create the 3D volume. The
  three curves corresponding to the lesser curvature, greater curvature and
  the center line are visible. Also visible are the elliptical cross sections
  used to generate the stomach profile at distinct points along the
  centerline. The image in the middle shows the outer surface of the composite
  3D volume which consists of the stomach cavity and the circular muscle
  layer. The rightmost image shows a cross sectional slice of the final
  structure used for the circular muscle layer. The thickness is highlighted
  in red. The structure is hollow which will be occupied by the fluid during
  IB simulations. This 3D model was created using SolidWorks R2018.}
  \label{fig:stomach_3d_solidworks}
\end{figure}

It is clear from Fig.~\ref{fig:stomach_pictures} that the human stomach has
three primary structural features. The first being the rounded top known as
the fundus. The second is the curved nature of the organ which leads to an
upper and lower surface commonly known as the `lesser' and `greater'
curvature. Finally, there is a pyloric sphincter at the end of the stomach
which controls and restricts the flow of fluid into the duodenum and small
intestine for further digestion. Based on earlier studies, the cross sectional
area of the stomach is assumed to have an elliptical profile
\cite{Kou2018_bmmb, Gnerucci2020}. With these simplifications, we construct a
set of planes that follow the center line of the stomach profile as shown in
Fig.~\ref{fig:stomach_3d_solidworks}. This step is similar to the approach
taken by Ferrua et al. to construct a volume that depicts the stomach cavity \cite{Ferrua2010}.
In the present study, we wish to create the muscle walls that envelope this
cavity for our immersed structure. Each plane contains an elliptical cross
section that is constructed using the lesser and greater curvatures as
guidelines. Once the set of ellipses are available, they are used to create a
sweep that generates the volume that serves as the stomach's cavity. With a
similar operation, a larger volume is created that is the union of the cavity
and the inner circular muscle layer. With a boolean subtraction operation, the
cavity is excluded from the larger volume which results in the circular muscle
layer of the stomach wall. With a simple inflation operation, an outer thick
walled shell of this structure is created which serves as the longitudinal
muscle layer of the stomach. Table~\ref{tab:stomach_dimensions} summarizes the
dimensions of the structure used to describe the stomach model. With the
construction of these two structures, the required geometry for creating a
finite element mesh of the stomach's layers is complete.

\subsection{Assigning fiber directions for the circular and longitudinal muscle layers}

In addition to geometric details, one must also consider the specific
orientation of fibers within the stomach wall's muscle layers. During gastric
peristalsis, these fibers contract in a systematic way to induce a propagating
reduction in lumen area which then creates fluid flow within the stomach. As
seen in Fig.~\ref{fig:stomach_pictures}, the stomach has three identifiable
muscle layers: circular, longitudinal and oblique. They are named based on the
orientation of fibers in each layer with respect to the centerline. Unlike the
esophagus, these three stomach layers are not distinct from each other and
fibers from one layer can continue on into another layer in very complex ways.
For the purposes of this study, we model only the circular and longitudinal
muscle layers while acknowledging the fact that the oblique layer is equally
important and has a significant effect on gastric peristalsis during normal
function. In addition to these intricate muscle fiber orientations, the
stomach has `sling' fibers that have a significant effect on fluid entering
the stomach through the esophagus. As these fibers are part of the oblique
layer, they have not been implemented in this model. However their presence
must be accounted for to make accurate predictions from the computational
model.

\begin{figure}
  \centerline{\includegraphics[width=\textwidth]{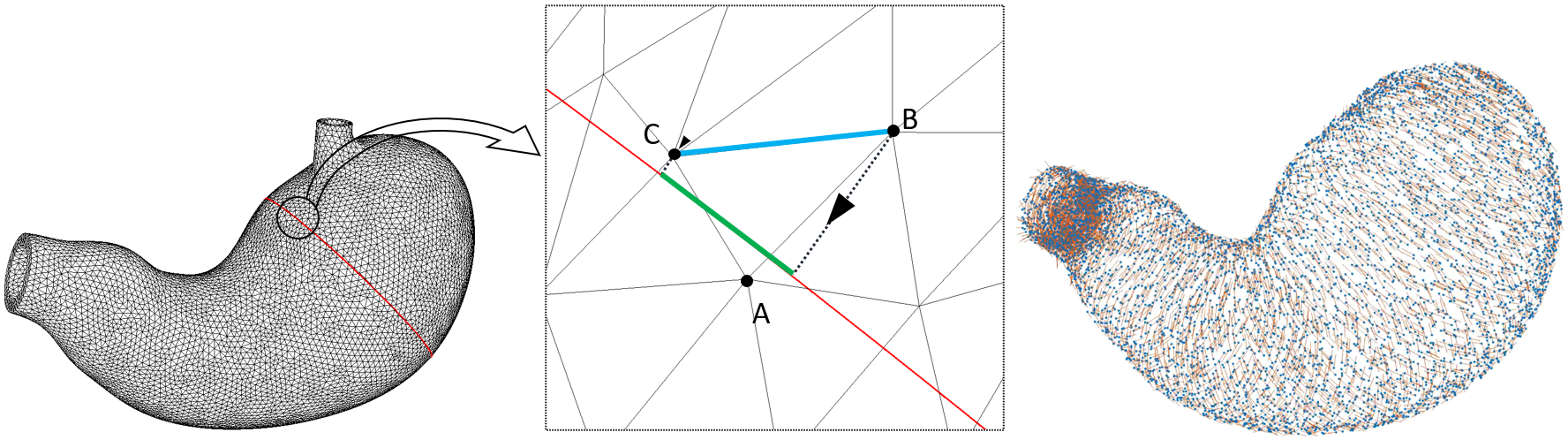}}
  \caption{A demonstration of assigning the circular muscle fiber orientation for an
  element using neighboring sides located on the solid boundary. The face ABC lies on
  the surface. Each of its sides are tangents to the solid surface. By
  projecting one of these tangents on to the plane, we obtain the fiber
  orientation for this element (shown in green). Image on the right shows the final
  fiber architecture for the circular muscle layer. Blue dots indicate element centers
  and brown lines indicate the element's fiber direction.}
  \label{fig:cm_layer_element_plane}
\end{figure}

To assign fiber directions for the circular muscle layer, we first generate a
finite element mesh of the structure using the open source meshing software
Gmsh \cite{Geuzaine2009}. This underlying mesh is used along with the planes
defined in Fig.~\ref{fig:stomach_3d_solidworks} to assign a fiber orientation
for each element using a simple procedure. For every element in the mesh, the
closest neightbor element with a face that lies on the solid's boundary is
found. As the circular muscle layer is relatively thin, such a neighbor is
relatively close to any element under consideration. The sides of this face
are oriented along the surface of the solid structure. A projection of one of
these sides on the plane closest to this element gives us the orientation of
the circular muscle fiber for this element. An example of this step is shown
in Fig.~\ref{fig:cm_layer_element_plane}. With the help of the boundary face,
we obtain vectors that are tangential to the surface. Projecting these tangent
vectors onto the planes that define the circular profiles gives us a vector
that is tangential to this circle. This procedure is repeated for
every element in the mesh which results in the final fiber architecture for
the circular muscle layer as seen in Fig.~\ref{fig:cm_layer_element_plane}.
During meshing, additional elements were added via local mesh refinement at
the pyloric sphincter region to better capture the small diameter profile and
its thicker walls. This leads to a greater density of fibers at the location
of the sphincter as is evident in the rightmost image of
Fig.~\ref{fig:cm_layer_element_plane}.

\begin{table}
\centering
\begin{tabular}{|l|c|c|}
\hline
Geometric feature                    & Value         & Source                              \\ \hline\hline
Length of lesser curvature           & 22.16 cm      & ---                                 \\ \hline
Length of greater curvature          & 33.25 cm      & Ferrua et al. \cite{Ferrua2011}     \\ \hline
Length of the centerline             & 24.8 cm       & ---                                 \\ \hline
Cavity volume                        & 0.85 L        & Ferrua et al. \cite{FERRUA20143664} \\ \hline
Widest diameter                      & 9.5 cm        & Schulze \cite{schulze2006}          \\ \hline
Pyloric sphincter diameter (relaxed) & 1.6 cm        & Deeg et al. \cite{Deeg2008}         \\ \hline
Muscle layer thickness (both LM and CM) & 2.75 mm        & Liebermann-Meffert et al. \cite{LiebermannMeffert1979}         \\ \hline
\end{tabular}
\caption{Dimensions of key features of the stomach model shown in
Fig.~\ref{fig:stomach_3d_solidworks}. When empty, stomach volume is around 250
mL \cite{Bouchard2010}. By choosing specific values for the lesser and greater curvature, 
cavity volume is increased to resemble the stomach geometry after the consumption of a
meal.}
\label{tab:stomach_dimensions}
\end{table}

\begin{figure}
  \centerline{\includegraphics[width=\textwidth]{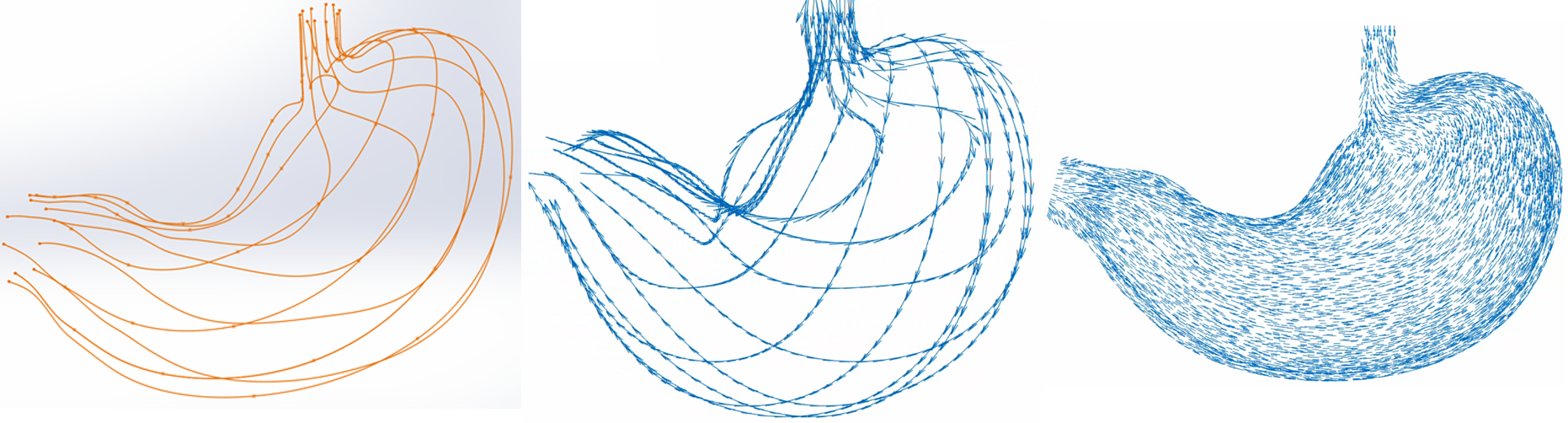}}
  \caption{Assigning fiber orientations to elements in the longitudinal muscle layer.
  Left: Guide curves drawn on the stomach surface. Center: Tangents to the
  guide curves used to assign fiber orientations to each element in the mesh.
  Right: Visualizing longitudinal fiber orientations for all elements in this
  muscle layer.}
  \label{fig:LM_guide_curves}
\end{figure}

For the longitudinal muscle layer fiber orientation, we first generate a set
of guide curves that travel along the surface of the stomach. Manually placed
points that lie on the outer surface of the stomach are joined using 3D
splines. Using each curve's trajectory, tangent
directions are computed for several equally spaced points on these curves as
seen in Fig.~\ref{fig:LM_guide_curves}. Following this step, for each element
center, we identify two curves that are closest to it. Points on these curves
that are closest to the element center are found and the average of the
tangent vector for each of these points is assigned as the fiber orientation
for said element. This ensures that fibers are generally oriented along the
trajectories of the guide curves and there is a smooth variation of fiber
vectors across all the elements in the layer. This process of assigning
longitudinal fiber directions and the final fiber orientations obtained is
visualized in Fig.~\ref{fig:LM_guide_curves}.

\subsection{Strain energy functions for the matrix and fiber components of the stomach wall}
\label{sec:LM_CM_sef_info}

In this section, we summarize the constitutive relations and material
properties used to describe the matrix and fiber components for the two muscle
layers of the stomach wall. Like most continuum mechanics-based models of
muscular tissue, each layer is assumed to be similar to a composite material
that consists of anisotropic fibers embedded in an isotropic tissue matrix.
Based on previous studies of esophageal transport by Kou et al.
\cite{Kou2017_jcp}, we use the bi-linear constitutive model proposed by Yang
et al. \cite{Yang2006} to describe the behavior of the circular and
longitudinal muscle layers of the gastric wall. Material constants pertaining
to the gastric wall are similar to the values chosen for the esophageal wall
in \cite{Kou2017_jcp}. It should be noted that there are significant
differences in material properties of the esophagus' circular and longitudinal
muscle layers compared to the corresponding muscle layers of the stomach wall.
However, as fibers in the stomach originate from respective muscle layers in
the esophagus \cite{smith_netter_2011}, we choose the same material properties
as a first order approximation of the system. The strain energy function
$\psi$ for each layer of the stomach structure is thus written the sum of two
parts
\begin{equation}
\psi = \psi_\mathrm{matrix}	+ \psi_\mathrm{fiber},
\end{equation}
which uses the incompressible neo-Hookean constitutive model for the isotropic
matrix and a bi-linear fiber model with orientation $\mathbf{a}$ for
individual fibers. The orientation vector $\mathbf{a}$ is a function of
$\mathbf{s}$ and its distribution was summarized in
Figs.~\ref{fig:cm_layer_element_plane} and \ref{fig:LM_guide_curves}. The
complete expression for $\psi$ used for the circular (CM) and longitudinal
(LM) layer, respectively can be written as
\begin{equation} \label{eq:CM_LM_sef}
\psi_\mathrm{CM/LM} = \psi_\mathrm{matrix}	+ \psi_\mathrm{fiber} = 
\frac{C_\mathrm{CM/LM}}{2}\left(I_1 - 3\right) + 
\frac{C^\prime_\mathrm{CM/LM}}{2}\left(\frac{\sqrt{I_4}}{\lambda_\mathrm{CM/LM}} - 1\right)^2.
\end{equation}
Here, the first invariant of the right Cauchy-Green strain tensor $\mathbb{C}$
is denoted as $I_1$ and the magnitude of stretch of an individual fiber is
denoted as $I_4 = \mathbb{C}:(\mathbf{a}\otimes\mathbf{a})$. The
non-dimensional rest length of a CM or LM fiber is denoted by
$\lambda_\mathrm{CM/LM}$ and its default value is 1.0, indicating that the
reference configuration of the element is equivalent to its undeformed shape.
Muscular contractions are generated by systematically reducing these rest
lengths to a value less than 1. The element then contracts to achieve a zero
stress-state, thus reducing the length of the fiber as is observed during a
muscular contraction. The following values were chosen for the material
constants occurring in Eqn.~\ref{eq:CM_LM_sef}: $C_\mathrm{CM/LM} = 0.4\
\mathrm{kPa},\ \mathrm{and}\ C^\prime_\mathrm{CM/LM} = 4.0\ \mathrm{kPa}$.

\section{Application of gastroesophageal modeling principles to simulate commonly observed physiological processes}

In the previous sections, we outlined the formulation and construction of the
building blocks needed to model physiological processes in the upper
gastrointestinal tract. In this section, we systematically assemble these
models to showcase three examples of gastrointestinal motility and fluid flow
observed due to the motion of the esophagus and stomach's walls.

\subsection{Example 1: Simulation of flow and wall motion during gastric peristalsis}

In the first application of the concepts outlined above, we present muscular
contraction-induced gastric peristalsis in a closed stomach. Unlike previous
studies, gastric peristalsis is induced not by specifying time-varying nodal
positions of the stomach wall, but by activating muscle fibers within the
elastic structure to reduce lumen area. As explained in
Sec.~\ref{sec:LM_CM_sef_info}, contractions can be induced by locally reducing
the value of $\lambda$. By varying this value along the stomach walls in a
controlled manner, we can generate a contraction that travels from the body of
the stomach, towards the pylorus. For this example, we assume that the
stomach is closed and is filled with a homogeneous fluid similar to water but
with a higher viscosity $(\mu = 0.01\ \mathrm{Pa\cdot s})$. These fluid
properties are identical to the ones used to study the transport of a bolus
from the esophagus into the stomach in Kou et al. \cite{Kou2018_bmmb}.

Due to the complexity of the stomach's geometry, it is not possible to use a
straightforward mathematical expression to reduce $\lambda$ in a wave-like
manner to induce peristalsis. To achieve peristalsis, we `activate' the planes
defined in Fig.~\ref{fig:stomach_3d_solidworks} in a sequential manner to
contract elements that are near specific planes. First, we activate the plane
that spans the thickest part of the pyloric sphincter. This ensures that the
sphincter is closed and does not allow any fluid to leave the cavity. This is
similar to the normal functioning of the stomach where the pylorus offers a
very high resistance to flow so that mixing is optimum before fluid enters the
intestine \cite{Avvari2021}. Subsequently, each plane starting from the body
of the stomach is activated one after the other. When a plane is activated,
all elements near it have their rest lengths changed using the following expression:
\begin{equation}
r_{\mathrm{curr}} = 
r_{\mathrm{max}}\exp\left(-\frac{1}{2\sigma^{2}}\left(\frac{t-t_{\mathrm{avg}}}{t_{b}-t_{a}}\right)^{2}\right)	
\end{equation}
Here, $(1 - r_\mathrm{curr})$ is the activated value of $\lambda$ and $(1 -
r_\mathrm{max})$ is the smallest value of $\lambda$ which occurs when the
contraction strength is at its peak at any given location. Each plane is
active for time interval $ t_a \leq t \leq t_b$ and 
$t_\mathrm{avg}=(t_a+t_b)/2$. The exponential function with $\sigma=1$ ensures
that elements near inactive planes have $r_\mathrm{curr}\approx 0$ and as $t$
exceeds $t_b$, $r_\mathrm{curr}$ gradually goes back to zero. This ensures
that the contraction travels smoothly along the stomach. It should be noted
that the contraction intensity of all elements activated by each plane is
uniform and is not a function of distance of the element from the plane.

\begin{table}[]
\centering
\begin{tabular}{|c|c|c|c|c|}
\hline
Plane num. & \begin{tabular}[c]{@{}c@{}}Activation\\ begin time $(t_a)$\end{tabular} & \begin{tabular}[c]{@{}c@{}}Activation\\ end time $(t_b)$\end{tabular} & \begin{tabular}[c]{@{}c@{}}Max. contraction\\ strength ($r_\mathrm{max}$)\end{tabular} & \begin{tabular}[c]{@{}c@{}}Plane\\ status info\end{tabular}                                    \\ \hline
1          & ---                                                             & ---                                                           & 0.0                                                                    & always inactive                                                                                \\ \hline
2          & ---                                                             & ---                                                           & 0.0                                                                    & always inactive                                                                                \\ \hline
3          & ---                                                             & ---                                                           & 0.0                                                                    & always inactive                                                                                \\ \hline
4          & 1.0                                                             & 1.5                                                           & 0.2                                                                    & ---                                                                                            \\ \hline
5          & 1.25                                                            & 1.75                                                          & 0.3                                                                    & ---                                                                                            \\ \hline
6          & 1.5                                                             & 2.0                                                           & 0.4                                                                    & ---                                                                                            \\ \hline
7          & 1.75                                                            & 2.25                                                          & 0.5                                                                    & ---                                                                                            \\ \hline
8          & 2.0                                                             & 2.5                                                           & 0.5                                                                    & ---                                                                                            \\ \hline
9          & 2.25                                                            & 2.75                                                          & 0.5                                                                    & ---                                                                                            \\ \hline
10         & 2.5                                                             & 3.0                                                           & 0.5                                                                    & ---                                                                                            \\ \hline
11         & 2.75                                                            & 3.25                                                          & 0.5                                                                    & ---                                                                                            \\ \hline
12         & 0.0                                                             & ---                                                           & 0.2                                                                    & \begin{tabular}[c]{@{}c@{}}pylorus plane\\ (permanently active\\ after t = 1.0 s)\end{tabular} \\ \hline
13         & ---                                                             & ---                                                           & 0.0                                                                    & always inactive                                                                                \\ \hline
14         & ---                                                             & ---                                                           & 0.0                                                                    & always inactive                                                                                \\ \hline
\end{tabular}
\caption{Complete set of activation timings and contraction strengths for each stomach plane used to simulate gastric peristalsis.}
\label{tab:gastric_peristalsis_ta_tb}
\end{table}

\begin{figure}[t!]
    \centering
    \begin{subfigure}[b]{0.3\textwidth}
        \centering
        \includegraphics[width=0.95\textwidth]{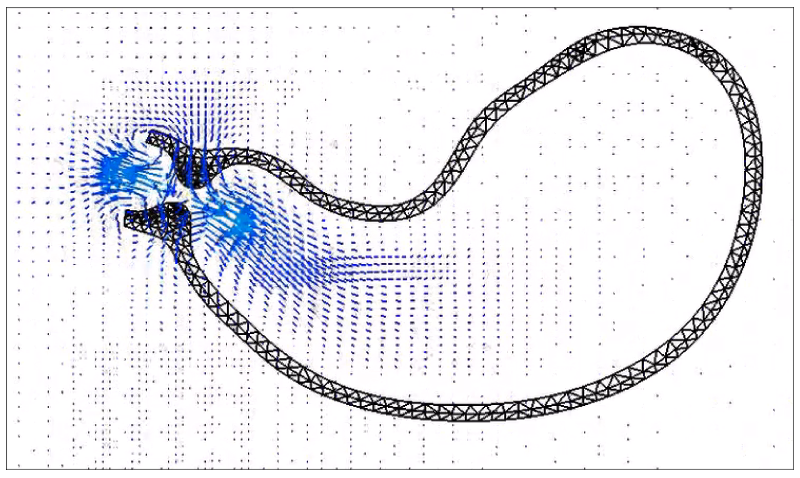}
        \caption{Pylorus closed and velocity fields near it.}
        \label{fig:gastric_peristalsis_closed_pylorus}
    \end{subfigure}
    \quad
    \begin{subfigure}[b]{0.3\textwidth}
        \centering
        \includegraphics[width=0.95\textwidth]{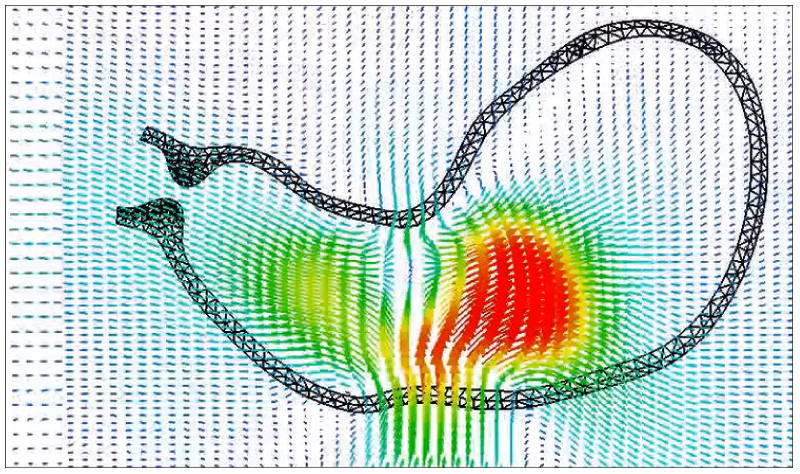}
        \caption{Contraction begins in the stomach body.}
        \label{fig:gastric_peristalsis_body_start}
    \end{subfigure}
    \begin{subfigure}[b]{0.3\textwidth}
        \centering
        \includegraphics[width=0.95\textwidth]{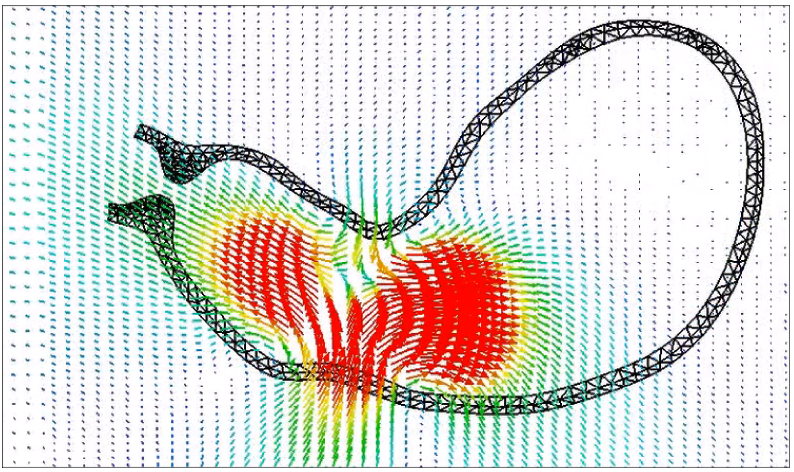}
        \caption{Contraction continues along the stomach.}
        \label{fig:gastric_peristalsis_body_1}
    \end{subfigure}
    \vskip\baselineskip
    \begin{subfigure}[b]{0.3\textwidth}
        \centering
        \includegraphics[width=0.95\textwidth]{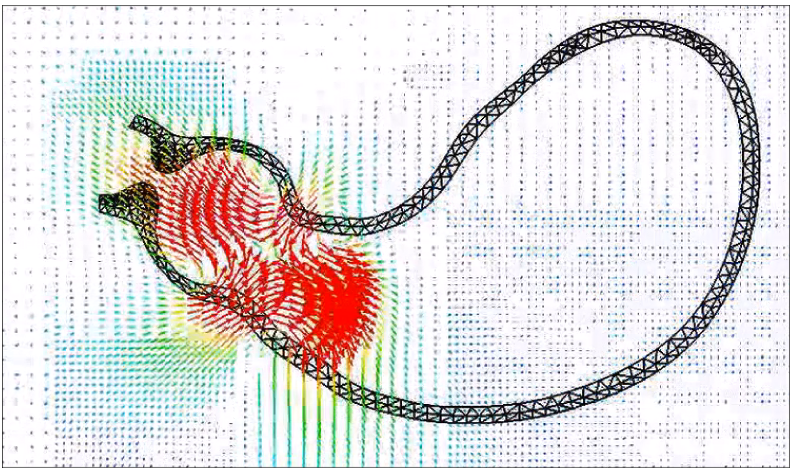}
        \caption{Contraction at the midsection.}
        \label{fig:gastric_peristalsis_body_mid}
    \end{subfigure}
    \quad
    \begin{subfigure}[b]{0.3\textwidth}
        \centering
        \includegraphics[width=0.95\textwidth]{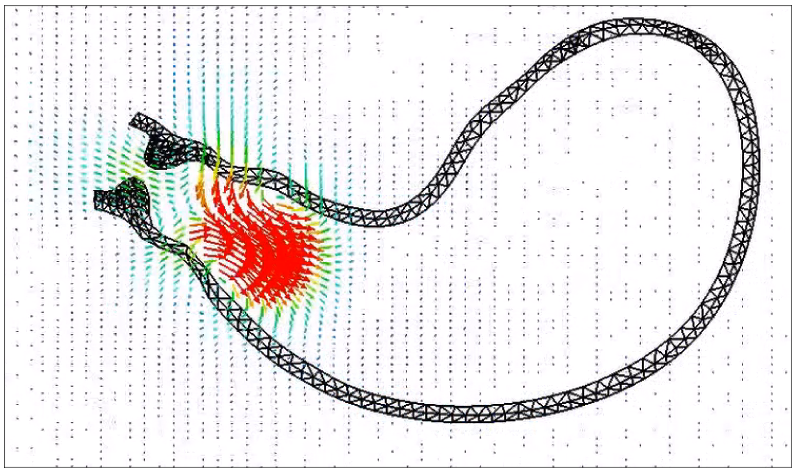}
        \caption{Contraction near the pylorus.}
        \label{fig:gastric_peristalsis_near_pylorus}
    \end{subfigure}
    \quad
    \begin{subfigure}[b]{0.3\textwidth}
        \centering
        \includegraphics[width=0.95\textwidth]{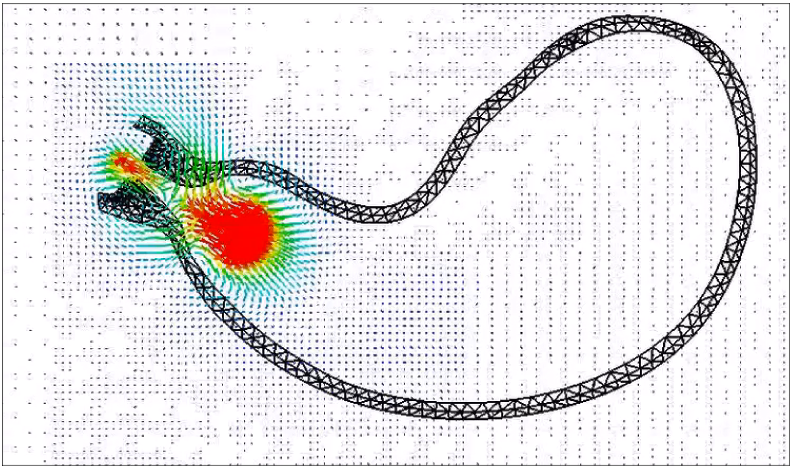}
        \caption{Contraction ends.}
        \label{fig:gastric_peristalsis_ends}
    \end{subfigure}
    \caption{Velocity vector fields observed during gastric peristalsis. Also seen is the generation of a fluid jet due to the contraction wave. Velocity vectors in the first frame show fluid flow around the pylorus as it is being closed.}
    \label{fig:gastric_peristalsis_velocity_shape}
\end{figure}

Velocity fields and the resulting stomach geometry observed during gastric
peristalsis is shown in Fig.~\ref{fig:gastric_peristalsis_velocity_shape}.
With the pylorus partially closed, circular muscle contractions force fluid
towards the distal part of the stomach. Due to the contraction acting on a
confined fluid, fluid is forced to travel in a retrograde direction with
respect to the traveling peristaltic wave. This formation of a `jet' is well
reported in numerical and clinical studies of digestion  in the stomach and
our model reproduces this behavior as well. Although the rest length of fibers
in the pyloric sphincter are reduced to induce closure, the pyloric sphincter
is not fully closed. This leads to a small amount of fluid to exit the stomach
cavity during peristalsis. This is similar to emptying in the stomach where a
majority of the contents are forced back towards the proximal region and a
small volume of fluid leaves to enter the duodenum.

\subsection{Example 2: Simulation of gravity-driven bolus emptying}

Ingested food enters the stomach by traveling along the esophageal lumen and
across the esophagogastric junction (EGJ). In the previous sections, we
outlined the construction of the stomach geometry and simulation of gastric
peristalsis. In this section, we combine this model with our previously
developed model for the esophagus \cite{Kou2017_jcp,Kou2018_bmmb} to simulate
emptying due to gravity. Prior studies did not account for the effect of
gravity and the difference in density between ingested bolus and the
surrounding fluid. In this example, we introduce a pill-shaped bolus (i.e., a
cylinder capped at both ends resembling a capsule) with a greater density and
viscosity compared to the surrounding fluid. The initial level-set function
used to generate the pill-shape is as follows,
\begin{equation} \label{eq:clamp_fn}
	h = \mathrm{min} \left\{ \mathrm{max}\left(\frac{\left(\mathbf{p}-\mathbf{a}\right)\cdot\left(\mathbf{b}-\mathbf{a}\right)}{\left(\mathbf{b}-\mathbf{a}\right)\cdot\left(\mathbf{b}-\mathbf{a}\right)},0.0\right),1.0 \right\}
\end{equation}

\begin{equation} \label{eq:sdf_calc}
	\mathrm{sdf}(\mathbf{p}) = \| \left(\mathbf{p}-\mathbf{a}\right) - \left(\mathbf{b}-\mathbf{a}\right)h\| - r.
\end{equation}
Here, $\mathbf{p}$ represents the three dimensional coordinates of any point
in the domain whose signed distance from the capsule surface needs to be
computed. The three-dimensional vectors $\mathbf{a}$ and $\mathbf{b}$ denote
the centers of the circles that form the ends of the cylindrical portion of
the pill shape. The radius of the cylinder and the spherical volumes that
cap the cylinder at both ends are equal and denoted by $r$. Equation
\ref{eq:clamp_fn} \textit{clamps} the value of $h$ between 0 and 1 depending
on the location of $\mathbf{p}$ relative to the ends of the cylinder. This
value is then used to compute the signed distance function for all points in
the domain using Eq. \ref{eq:sdf_calc}. All points within the capsule region
(with a negative level set value) were assigned a density of $1.5\rho_f$ and a
 dynamic viscosity of $3\mu_f$ where $\rho_f$ and $\mu_f$ are material
properties of the surrounding ambient fluid taken from previous studies
\cite{Kou2017_jcp}. This formed the initial shape of the bolus with a volume
of 3.5 mL, occupying the esophageal lumen before entering the stomach.

\begin{figure}[t]
  \centerline{\includegraphics[width=\textwidth]{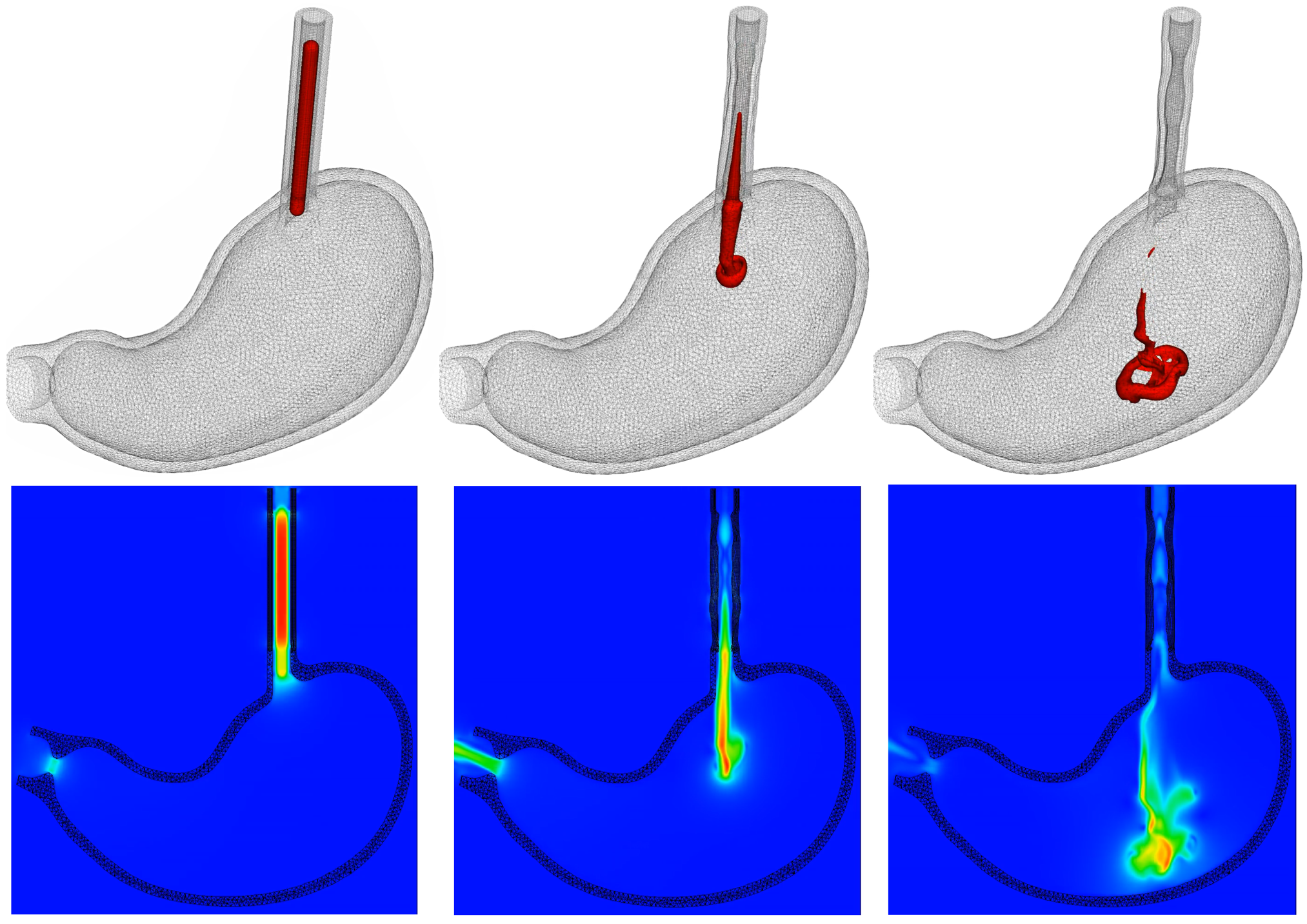}}
  \caption{Gravity-driven bolus emptying into the stomach. Figures in the top row show
  the evolution of the surface of the bolus as it enters the stomach. This is
  constructed by visualizing the zero-level set surface in the domain. Figures
  in the bottom row show the variation of fluid velocity in a two-dimensional
  plane slicing the esophagus and stomach in half. Velocity scales are
  modified for each figure to ensure fluid motion in the domain is clearly
  visible at every instant.}
  \label{fig:stomach_bolus_emptying_sim}
\end{figure}

The evolution of the bolus as it enters the stomach is shown in
Fig.~\ref{fig:stomach_bolus_emptying_sim}. The first instant of
Fig.~\ref{fig:stomach_bolus_emptying_sim} shows the location of the zero-level
set as constructed by Eqn.~\ref{eq:sdf_calc}. For this test case, the active
portion of the stomach wall is disabled. This is based on the fact that
gastric peristalsis is rarely observed when the volume of fluid entering the
stomach is small, as it is specified for this test case. The bolus fully exits
the esophgus while the peristaltic contraction just begins traveling over the
esophagus. This shows that the difference in density between the bolus and the
ambient fluid was sufficient to achieve gravity-induced bolus transport. It
must be emphasized that there is a subtle but important difference between the
simulation shown here and emptying occurring in a human subject. When food
enters the esophagus, the stomach walls expand, allowing its volume to
increase in a process called \textit{gastric accommodation}. This leads to a
decrease in pressure in the gastric cavity allowing the bolus to enter the
stomach . In this test case however, emptying is made possible by leaving the
pyloric sphincter open. One can observe the development of greater fluid
velocity at this location as the bolus enters the stomach. Due to ambient
fluid exiting the stomach cavity through the pylorus, the heavier bolus is able to enter to
stomach and break up as shown in Fig.~\ref{fig:stomach_bolus_emptying_sim}. We
address this limitation below and future work will be geared towards
implementing realistic behavior of the stomach walls to model gastric
accommodation.

\subsection{Example 3: Simulating a transient LES relaxation (tLESR) event and 
retrograde flow of stomach contents into the esophagus}

In this final example showcasing the application of the concepts presented
previously, we construct a test case for retrograde (reverse) flow of stomach
contents into the esophagus, driven by density-driven buoyancy forces. Similar
flow patterns occur during commonly observed phenomena like eructation
(belching), acid reflux and emesis (vomiting). Under normal conditions, the
EGJ remains closed and prevents acidic stomach contents from entering the
esophagus and harming its delicate inner mucosal layer. Retrograde flow occurs
when this barrier is opened either by infrequent muscular relaxation in a
healthy subject or mechanical weakening due to pathological developments like
hiatal hernia or a hypotensive LES \cite{Gor2015}. There are two primary
forces that drive this retrograde flow: 1) greater pressure in the gastric
cavity due to contraction or tone in the walls of the fundus, or, 2)
differences in density between the confined and ambient fluids, leading to a
pressure gradient generated by the buoyancy forces. For this test case, we
disable the active component in the stomach walls and consider it to be a
simple, hyperelastic container that serves to confine the fluid within its
boundaries. Thus, gravitational effects combined with the differences in
density lead to retrograde fluid flow.

\begin{figure}[t]
  \centerline{\includegraphics[width=\textwidth]{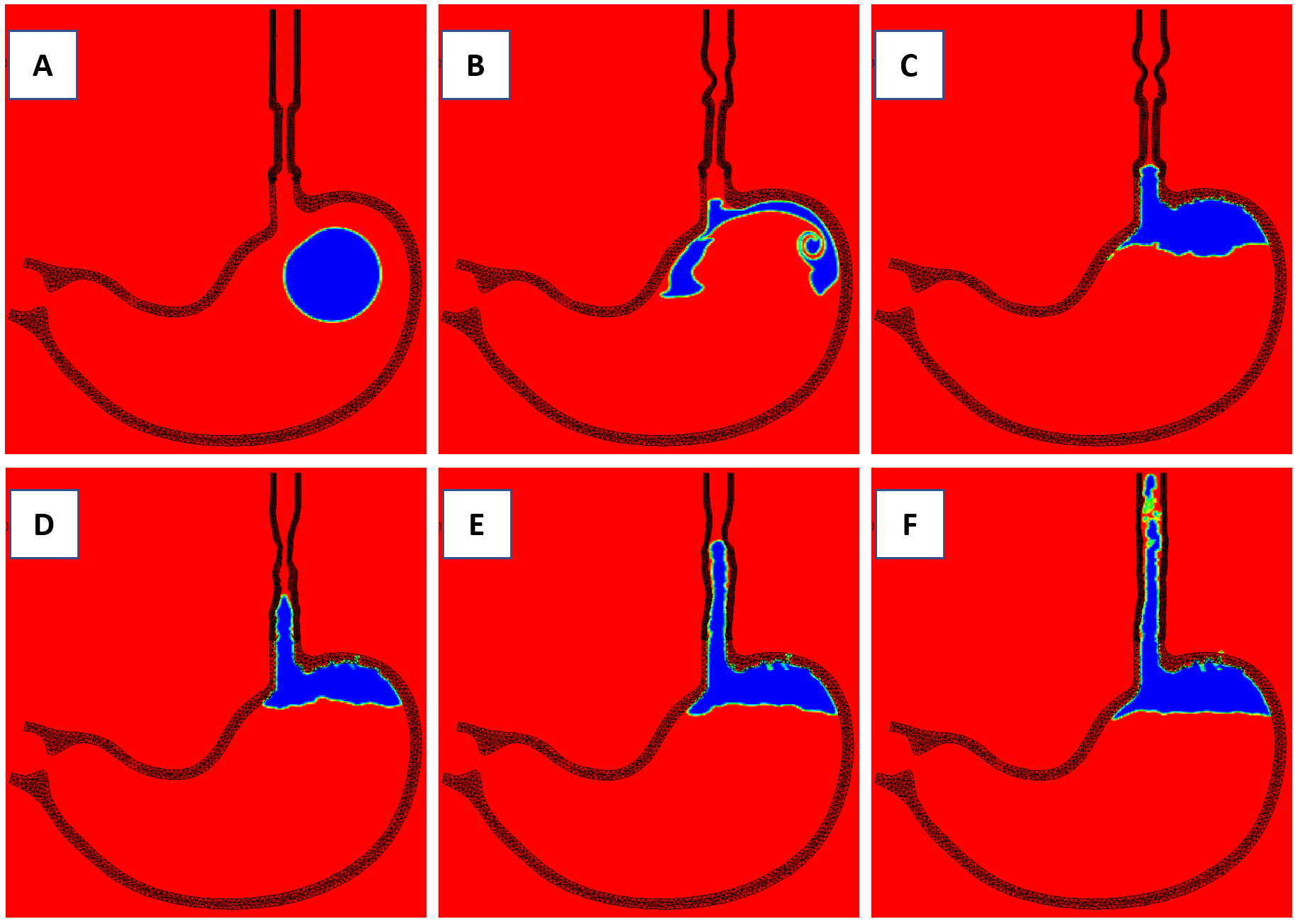}}
  \caption{Retrograde flow of a lower density fluid $(\rho=900\ \mathrm{kg/m}^3)$ from
  the stomach into the esophagus. Ambient fluid is water $(\rho=1000\
  \mathrm{kg/m}^3)$, shown in red. Figure (c) is a temporary steady state
  where the fluid has settled in the fundus and prevented from entering the
  esophagus by the closed sphincter. Figure (d) shows the density field
  shortly after the sphincter is programmed to open.}
  \label{fig:retrograde_flow_sim_imgs}
\end{figure}

Modeling this retrograde flow requires three key ingredients: (1) A closed
structural barrier that resembles a sphincter at the distal end of the
esophagus, (2) A lower density fluid resting against this closed barrier, and
(3) a controlled opening of the sphincter to allow this fluid to escape into
the esophagus. We obtain this configuration by first closing the sphincter as
shown in Fig.~\ref{fig:retrograde_flow_sim_imgs}a. The initial configuration
of the fluid is that of a spherical bubble in the middle of the gastric
cavity. The bubble is then allowed to rise and reach the closed sphincter. The
fluid is then allowed to reach steady state as shown in
Fig.~\ref{fig:retrograde_flow_sim_imgs}c. After this steady state
configuration is reached, the sphincter opening is modeled by increasing the
rest length of the circular muscle fibers in the closed segment. A similar
behavior occurs during transient lower esophageal sphincter relaxation
(tLESR).  This spontaneous relaxation of the LES periodically allows air
trapped in the stomach to escape \cite{Kim2013}. Following this type of
opening in the model presented here, the lighter fluid is then able to
increase the area of the sphincter and enter the esophageal lumen as shown in
Fig.~\ref{fig:retrograde_flow_sim_imgs}f. Thus, this model is able to account
for all three features which are necessary to develop future models to study
tLESRs, acid reflux and the mechanical competancy of the esophagogastric
junction to prevent reflux.

There are several key things to note about the succesful implementation of
this model. Observe that the LES does not appear to be fully closed (no
physical contact between the walls as seen in
Fig.~\ref{fig:retrograde_flow_sim_imgs}a). Due to the delta function kernels
used in the immersed boundary method, the fluid-solid interface is smeared.
This ensures that although the LES does not \textit{appear} to be closed, it
is \textit{effectively} closed because the Lagrangian forcing from the
esophagus' walls extends into the lumen and offers some resistance to fluid
flow. Another physiological detail that must be emphasized is that the
distance between the LES and the stomach was increased to make the
physiological problem more amenable for simulation. In reality, the LES spans
the region where the stomach meets the cylindrical tube forming the angle of
His. Similarly, the effect of surface tension between the two fluids has not
been modeled. Finally, the most important limitation of this model was the
small difference in density between the ambient and trapped fluid in the
stomach. Ideally, the density ratio modeled should be equal to $10^3$,
reflecting a problem involving air and water. However, this led to significant
buoyancy forces being applied on the entire stomach structure causing large,
unrealistic deformations. Thus, a smaller density ratio was chosen to develop
this test case. This deficiency can be rectified by including the restraining
effect of the diaphragm on the stomach and ensuring that the esophagus is
filled with air as it is in human subjects. Future modeling attempts will
account for these details and treat the contents of the stomach as air and
water with the appropriate surface tension effects for a better depiction of
the true physical problem.

\section{Model Limitations and Future Work}

Due to the immense complexity of any biological organ system coupled with
limitations on computational resources available, it is difficult to expect
a numerical model to capture all the physics involved in normal physiologic
function. For the gastrointestinal tract in particular, it is not only
homogeneous fluid and solid mechanics, but also granular flow, chemical and
tissue electro-mechanical activity that governs overall organ function. The
model presented in this work does not account for the absorption and
mechanical breakdown of solid food in the stomach due to the pulverizing
action of gastric peristalsis. It also does not account for the complex
interaction between the structure's strain fields and their influence on
intensity of peristalsis and gastric accommodation which normally occurs in
the stomach walls as they distend and respond to the amount of food due to
distension-induced mechanoreceptors \cite{Takeda2004, Barlow2002}. As
explained in Sec.~\ref{sec:intro}, models presented by other researchers have
accounted for electrical activity or chemical kinetics but no work has been
done that accounts for the most important physical interaction in the GI
system i.e. between the fluid contents and the elastic structures that modify
their flow fields.

Another important but rectifiable limitation of this work is the usage of an
idealized stomach and esophagus geometry. The assigned fiber-architecture was
also procedurally generated and not based on real clinical imaging. In
reality, the esophagus is not a perfectly straight cylinder and the stomach's
shape is far more complex than shown in Fig.~\ref{fig:stomach_3d_solidworks}.
For the purposes of this study, we chose to construct the 3D model from
pictorial representations of the stomach in commonly available literature on
human anatomy. Muscle layers were then constructed based on rough estimates of
muscle thickness and inflating the idealized stomach cavity volume. The
oblique muscle layer was not modeled as well. Similar to the esophagus, the
stomach has mucosal and submucosal layers with a large amount of gastric
folds. These can unfold when the stomach expands to increase its volume. In
addition to this, these folds can affect the fluid's velocity profile near the
stomach wall. Thus, assuming a smooth inner surface for the stomach wall is
another limitation of this study. Ideally, the esophagus and stomach geometry
would be obtained from medical imaging of a subject (e.g., MRI or CT scan) and
segmenting the data to isolate various muscle layers and generate a realistic
3D structure for simulation. In a future work, we aim to obtain data from
4D-MRI scans \cite{Markl_4DMRI_2014} and generate patient-specific geometries
for improved simulation and analysis. For muscle fiber architecture, diffusion
tensor imaging \cite{Liao2018, Gilbert2008} or sectioning microscopy
\cite{Zifan2017, Yassi2010} has been used to find the orientation of
individual fibers. However, for future studies, we aim to stick to rule-based
algorithms \cite{Bayer2012} to assign fiber architecture and move to medical
imaging-based approaches when some of the more pressing limitations have been
addressed.

In addition to the technical limitations mentioned above, it must be noted
that there were a few features of the esophagogastric junction (EGJ) that were
not fully captured in the present model. The EGJ is the location where the
esophagus continues on into the stomach. It is an anatomically complex region
with two important components: 1) an intrinsic sphincter, commonly known as
the lower esophageal sphincter i.e. the LES, and, 2) an extrinsic sphincter
which consists of the crural diaphragm and other surrounding structures at the
EGJ \cite{Tack2018}. The combination of these two structures forms the primary
barrier that prevents stomach acids from entering the esophagus. In
the model presented above, the LES is adequately captured by the circular
muscle fiber architecture which can be programmed to contract or relax to
allow fluid to enter the stomach. However, the extrinsic sphincter which
consists of sling fibers from the crural diaphragm has not been modeled. These
fibers wrap around the esophagus and are responsible for creating a
significant asymmetry in the pressure profile at the EGJ \cite{Mittal2017}. In
addition to these fibers, an additional structure called the phrenoesophageal
ligament (PEL) is responsible for tethering the esophagus to the diaphragm.
This tethering adds a significant amount of longitudinal tension to the
esophagus which can affect the motion of the EGJ during emptying or reflux.
This tension is primarily responsible for the esophagus returning to its
original configuration after longitudinal muscle contraction
\cite{Bombeck1966}. The presence of these two elements, i.e. the PEL and sling
fibers, is particularly important in the context of gastroesophageal reflux
disease (GERD) and hiatal hernia as their integrity is compromised in this
disease state. As such, future efforts will be acutely directed towards
developing a detailed model of the PEL and gastric sling fibers to properly
capture their overall effect on EGJ function in normal and pathological
scenarios.

\section*{Acknowledgments}

This research was supported in part through the computational resources and
staff contributions provided for the Quest high performance computing facility
at Northwestern University which is jointly supported by the Office of the
Provost, the Office for Research, and Northwestern University Information
Technology.

This work also used the Extreme Science and Engineering Discovery Environment
(XSEDE) clusters Comet, at the San Diego Supercomputer Center (SDSC) and
Bridges, at the Pittsburgh Supercomputing Center (PSC) through allocation
TG-ASC170023, which is supported by National Science Foundation grant number
ACI-1548562 \cite{xsede}.

\section*{Funding Data}
\begin{itemize}
\item National Institutes of Health (NIDDK grants DK079902 \& DK117824; Funder ID: 10.13039/100000062)
\item National Science Foundation (OAC grants 1450374 \& 1931372; Funder ID: 10.13039/100000105)
\end{itemize}

\bibliographystyle{elsarticle-num}
\bibliography{stomach_fsi_draft_references}

\end{document}